\input epsf

\magnification\magstephalf
\overfullrule 0pt
\def\gsim{\raise.3ex\hbox{$\;>$\kern-.75em\lower1ex\hbox{$\sim$}$\;$}}

\font\rfont=cmr10 at 10 true pt
\def\ref#1{$^{\hbox{\rfont {[#1]}}}$}


\font\fourteenbf=cmbx12 scaled\magstep1

\font\tenbfit=cmbxti10
\font\sevenbfit=cmbxti10 at 7pt
\font\fivebfit=cmbxti10 at 5pt
\newfam\bfitfam 
\textfont\bfitfam=\tenbfit  \scriptfont\bfitfam=\sevenbfit
\scriptscriptfont\bfitfam=\fivebfit

\font\tenbfit=cmbxti10
\font\sevenbfit=cmbxti10 at 7pt
\font\fivebfit=cmbxti10 at 5pt
\newfam\bfitfam 
\textfont\bfitfam=\tenbfit  \scriptfont\bfitfam=\sevenbfit
\scriptscriptfont\bfitfam=\fivebfit

\font\tenbit=cmmib10
\newfam\bitfam
\textfont\bitfam=\tenbit%

\font\tenmbf=cmbx10
\font\sevenmbf=cmbx7
\font\fivembf=cmbx5
\newfam\mbffam
\textfont\mbffam=\tenmbf \scriptfont\mbffam=\sevenmbf
\scriptscriptfont\mbffam=\fivembf

\font\tenbsy=cmbsy10
\newfam\bsyfam 
\textfont\bsyfam=\tenbsy%


\def\pmb#1{\setbox0=\hbox{#1}
 \kern.05em\copy0\kern-\wd0 \kern-.025em\raise.0433em\box0 }

\def\slash{/\kern-.5em}


 %


\def\boxit#1{\vbox{\hrule\hbox{\vrule\kern1pt\vbox
{\kern1pt#1\kern1pt}\kern1pt\vrule}\hrule}}

\def\h{\hfill\break}
\parskip=6pt
\parindent=0pt
\hsize=17truecm\hoffset=-5truemm
\vsize=23truecm
\def\footnoterule{\kern-3pt
\hrule width 17truecm \kern 2.6pt}


\catcode`\@=11 

\def\nolabels{\def\wrlabeL##1{}\def\eqlabeL##1{}\def\reflabeL##1{}}
\def\writelabels{\def\wrlabeL##1{\leavevmode\vadjust{\rlap{\smash%
{\line{{\escapechar=` \hfill\rlap{\sevenrm\hskip.03in\string##1}}}}}}}%
\def\eqlabeL##1{{\escapechar-1\rlap{\sevenrm\hskip.05in\string##1}}}%
\def\reflabeL##1{\noexpand\llap{\noexpand\sevenrm\string\string\string##1}}}
\nolabels
\global\newcount\refno \global\refno=1
\newwrite\rfile
\def\defref{$^{{\hbox{\rfont [\the\refno]}}}$\nref}
\def\nref#1{\xdef#1{\the\refno}\writedef{#1\leftbracket#1}%
\ifnum\refno=1\immediate\openout\rfile=refs.tmp\fi
\global\advance\refno by1\chardef\wfile=\rfile\immediate
\write\rfile{\noexpand\item{#1\ }\reflabeL{#1\hskip.31in}\pctsign}\findarg}
\def\findarg#1#{\begingroup\obeylines\newlinechar=`\^^M\pass@rg}
{\obeylines\gdef\pass@rg#1{\writ@line\relax #1^^M\hbox{}^^M}%
\gdef\writ@line#1^^M{\expandafter\toks0\expandafter{\striprel@x #1}%
\edef\next{\the\toks0}\ifx\next\em@rk\let\next=\endgroup\else\ifx\next\empty%
\else\immediate\write\wfile{\the\toks0}\fi\let\next=\writ@line\fi\next\relax}}
\def\striprel@x#1{} \def\em@rk{\hbox{}} 
\def\lref{\begingroup\obeylines\lr@f}
\def\lr@f#1#2{\gdef#1{\defref#1{#2}}\endgroup\unskip}
\newwrite\lfile
{\escapechar-1\xdef\pctsign{\string\%}\xdef\leftbracket{\string\{}
\xdef\rightbracket{\string\}}}

\def\writestop{\def\writestoppt{\immediate\write\lfile{\string\p
ageno%
\the\pageno\string\startrefs\leftbracket\the\refno\rightbracket%
\string\def\string\secsym\leftbracket\secsym\rightbracket%
\string\secno\the\secno\string\meqno\the\meqno}\immediate\closeout\lfile}}
\def\writestoppt{}\def\writedef#1{}
\catcode`\@=12 
\centerline{\fourteenbf Does the hard pomeron obey Regge factorisation?}
\vskip 8pt
\centerline{A Donnachie}
\centerline{Department of Physics, Manchester University}
\vskip 5pt
\centerline{P V Landshoff}
\centerline{DAMTP, Cambridge University$^*$}
\footnote{}{$^*$ email addresses: sandy.donnachie@man.ac.uk, \ pvl@damtp.cam.ac.uk}
\bigskip
{\bf Abstract}

While data for the proton structure function demand
the presence of a hard-pomeron contribution even at quite small $Q^2$, 
previous fits to the $pp$ and $p\bar p$ total cross sections have
found that in these there is little or no room for such a contribution.
We re-analyse the data and show that it 
may indeed be present and that, further, it probably obeys Regge
factorisation: 
$\sigma_{{\fiverm{\hbox{HARD}}}}^{\gamma p}(s,Q_1^2)~
\sigma_{{\fiverm{\hbox{HARD}}}}^{\gamma p}(s,Q_2^2)=
\sigma_{{\fiverm{\hbox{HARD}}}}^{pp}(s)~
\sigma_{{\fiverm{\hbox{HARD}}}}^{\gamma\gamma}(s,Q_1^2,Q_2^2)$
for all values of $Q_1^2$ and $Q_2^2$.

\vskip 8truemm
It has become traditional\defref\sigtot{
A Donnachie and P V Landshoff, Physics Letters B296 (1992) 227
}
to believe that soft hadronic processes at
high energy are dominated by the exchange of a soft pomeron with Regge
intercept close to $1+\epsilon_1=1.08$, or possibly a little 
larger\defref\cudell{
J R Cudell, K Kang and S K Kim, Physics Letters B395 (1997) 311
}. Data for the proton structure function $F_2(x,Q^2)$ have 
revealed\defref\twopom{
A Donnachie and P V Landshoff, Physics Letters B437 (1998) 408
and B518 (2001) 63
}
also the presence of a hard pomeron, with intercept $1+\epsilon_0$
a little greater
than 0.4. The hard pomeron is seen particularly clearly in the charm
structure function $F_2^c(x,Q^2)$, because experiment finds that,
while for $F_2(x,Q^2)$ both hard and soft pomeron contributions are
present, the soft pomeron does not contribute to $F_2^c(x,Q^2)$.

The data for $F_2(x,Q^2)$ and $F_2^c(x,Q^2)$ are consistent with the
hard pomeron being a simple Regge pole. 
One would therefore expect its contributions to different processes to
obey Regge factorisation\defref\book{
A Donnachie, H G Dosch, P V Landshoff and O Nachtmann, {\it Pomeron physics
and QCD}, Cambridge University Press (2002)
}. 
In particular, at each value of $W=\sqrt s$,
$$
\sigma_{{\fiverm{\hbox{HARD}}}}^{\gamma\gamma}(s,Q_1^2,Q_2^2)=
{\sigma_{{\fiverm{\hbox{HARD}}}}^{\gamma p}(s,Q_1^2)~
\sigma_{{\fiverm{\hbox{HARD}}}}^{\gamma p}(s,Q_2^2)\over
\sigma_{{\fiverm{\hbox{HARD}}}}^{pp}(s)}
\eqno(1)
$$
See figure 1.
These relations should hold for all values of $Q_1^2$ and $Q_2^2$,
both 0 and nonzero. For the case where one of them is 0, 
an equivalent statement is
$$
F_{2~{\fiverm{\hbox{HARD}}}}^{\gamma}(s,Q^2)=
{F_{2~{\fiverm{\hbox{HARD}}}}(x,Q^2)~
\sigma_{{\fiverm{\hbox{HARD}}}}^{\gamma p}(s)\over
\sigma_{{\fiverm{\hbox{HARD}}}}^{pp}(s)}
\eqno(2)
$$

Our previous strategy was first\ref{\sigtot} to fit data for purely-hadronic
total cross sections with only a soft pomeron and reggeon exchange,
that is $\rho,\omega,f_2,a_2$ exchange. This determined the intercept
of the soft pomeron to be close to $1+\epsilon_1=1.08$. 
We fixed this in our subsequent
fits\ref{\twopom} to the data for the proton structure function
$F_2(x,Q^2)$, and found that they are described very successfully
by adding in a hard pomeron with intercept a little larger than 
$1+\epsilon_0=1.4$. Keeping $\epsilon_1$ fixed, 
we then concluded there is little room to include a hard-pomeron term 
in the purely-hadronic cross sections.  We now adopt a different strategy,
and simultaneously fit data for $\sigma^{pp},\sigma^{p\bar p},\sigma^{\gamma p}$
and the proton structure function $F_2$, treating both $\epsilon_0$ and
$\epsilon_1$ as free parameters. 

As is well known, there is a conflict between the measurements of the
$p\bar p$ cross section at the Tevatron\defref\tevatron{
E710 collaboration: N Amos et al, Physical Review Letters 63 (1989) 2784;
CDF collaboration\h
 F Abe et al, Physical Review D50 (1994) 5550
}.
We assume that there is a better chance of
accommodating a hard pomeron in the $p\bar p$ total cross section if
we use the larger CDF measurement at the Tevatron, rather than E710.
The details of the fit are dependent on how far down in $\sqrt s$
we wish to extend it. We have chosen to go down to 6 GeV, where
previously we chose 10 GeV. After all, we do have extra parameters
when we introduce additional, hard-pomeron terms and so should be able to
achieve a better fit.
Although\ref{\book}
the intercepts of the $C=+1$ and $C=-1$ nonpomeron Regge trajectories
are not quite equal, we continue to use a single effective intercept
$1+\epsilon_R$ for them.

So our fits to total cross sections are of the form
$$
\sigma={1\over 2p\sqrt s}\Big(X_0(2\nu)^{1+\epsilon_0}
+X_1(2\nu)^{1+\epsilon_1} +Y(2\nu)^{1+\epsilon_R}\Big)
\eqno(3)
$$
For the proton structure function $F_2(x,Q^2)$ at small $x$ we use
$$
F_2(x,Q^2)\sim f_0(Q^2)x^{-\epsilon_0}+
f_1(Q^2)x^{-\epsilon_1}+f_R(Q^2)x^{-\epsilon_R}
\eqno(4a)
$$
with
$$
f_0(Q^2)=A_0 (Q^2)^{1+\epsilon_0}/(1+Q^2/Q_0^2)^{1+\epsilon_0/2}
$$$$
f_1(Q^2)=A_1 (Q^2)^{1+\epsilon_1}/(1+Q^2/Q_1^2)^{1+\epsilon_1}
$$$$
f_R(Q^2)=A_R (Q^2)^{1+\epsilon_R}/(1+Q^2/Q_R^2)^{1+\epsilon_R}
\eqno(4b)
$$

\topinsert
\centerline{\epsfxsize=0.6\hsize\epsfbox[0 0 590 290]{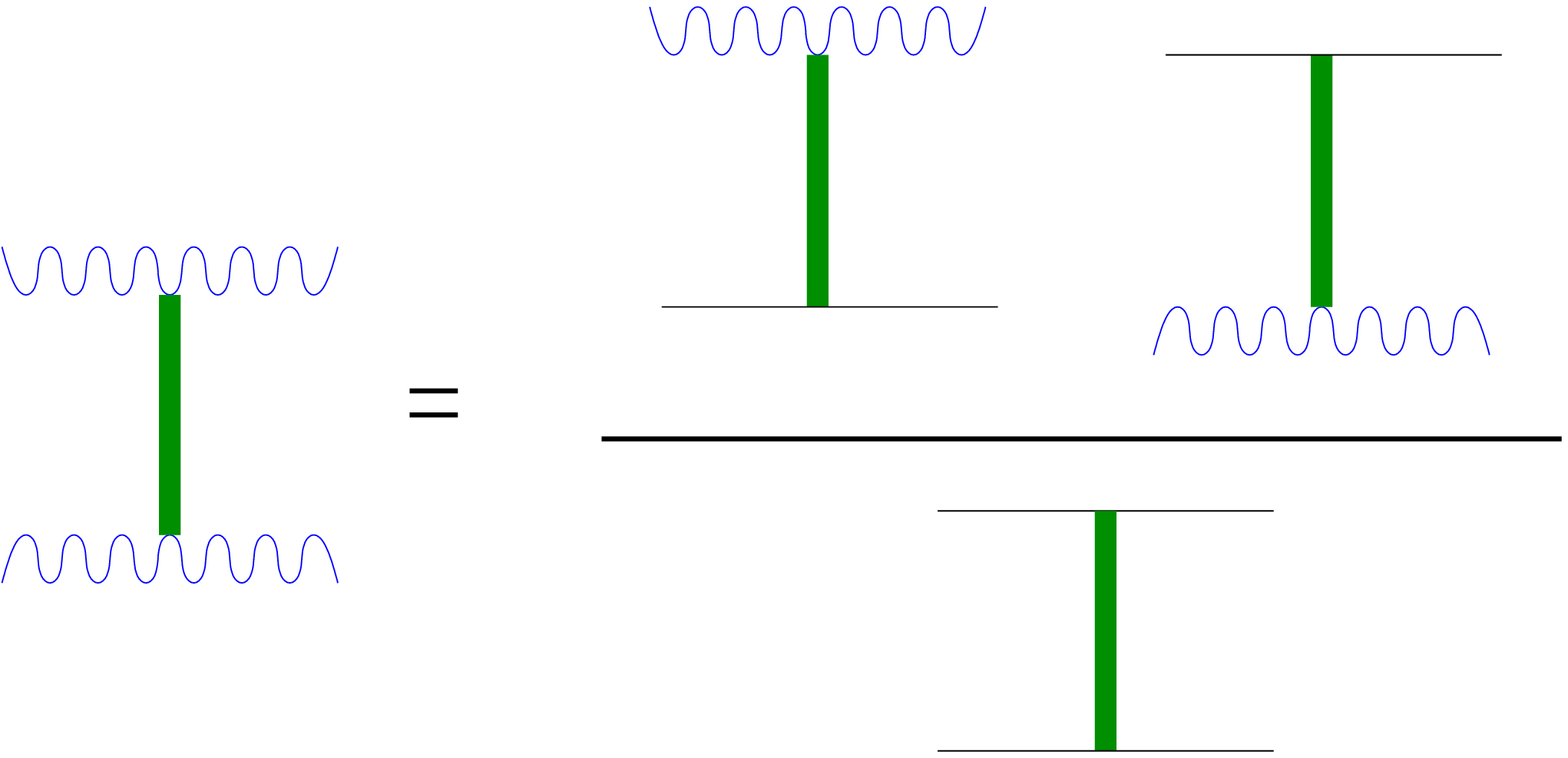}}

\centerline{Figure 1: Regge factorisation}
\endinsert

\topinsert
\line{
\epsfxsize=0.45\hsize\epsfbox[125 495 470 770]{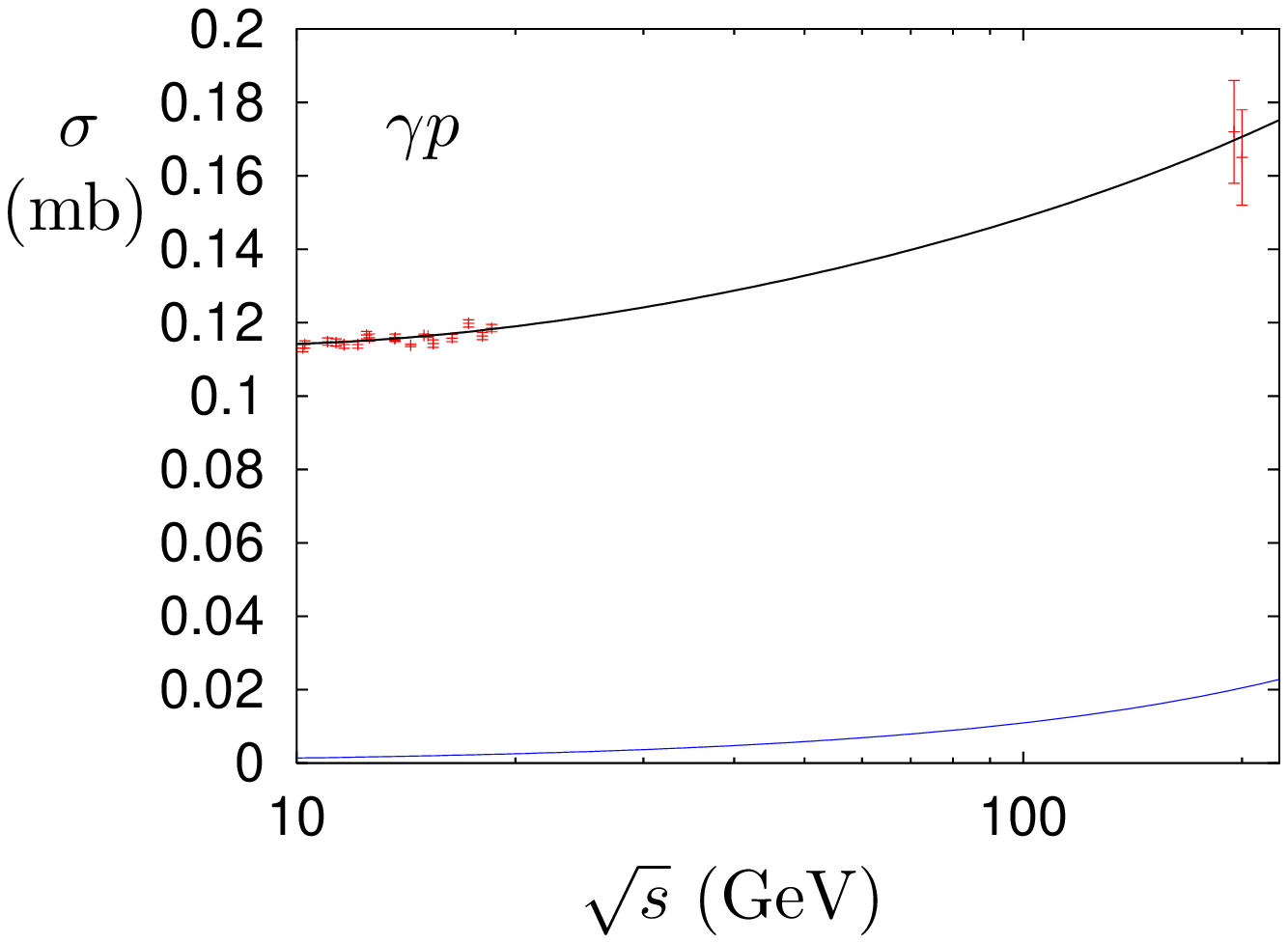}\hfill
\epsfxsize=0.45\hsize\epsfbox[125 495 470 770]{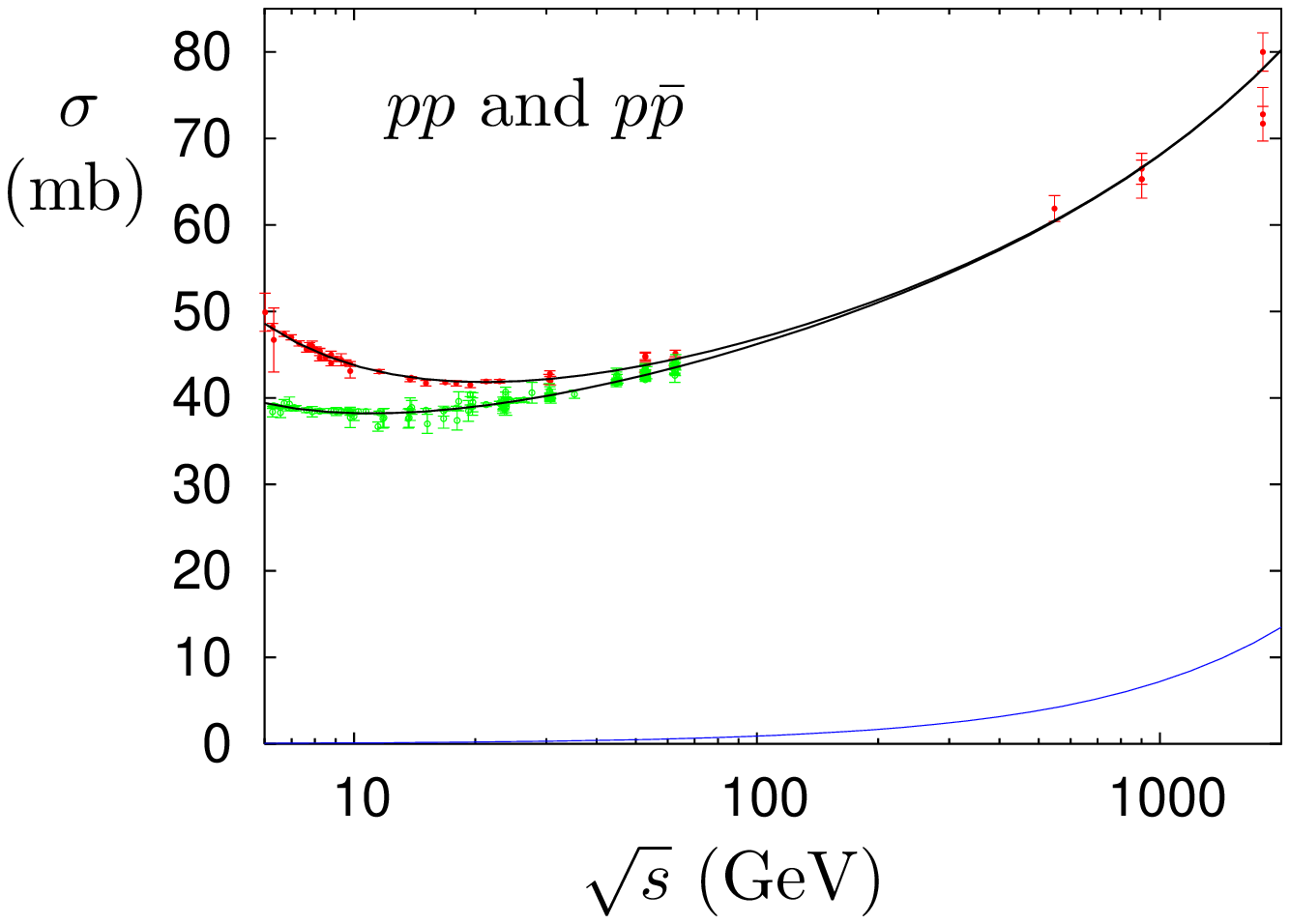}}

\line{\hfill (a)\hfill\hfill (b)\hfill}
\vskip 6truemm
\centerline{\epsfxsize=0.55\hsize\epsfbox[65 295 377 755]{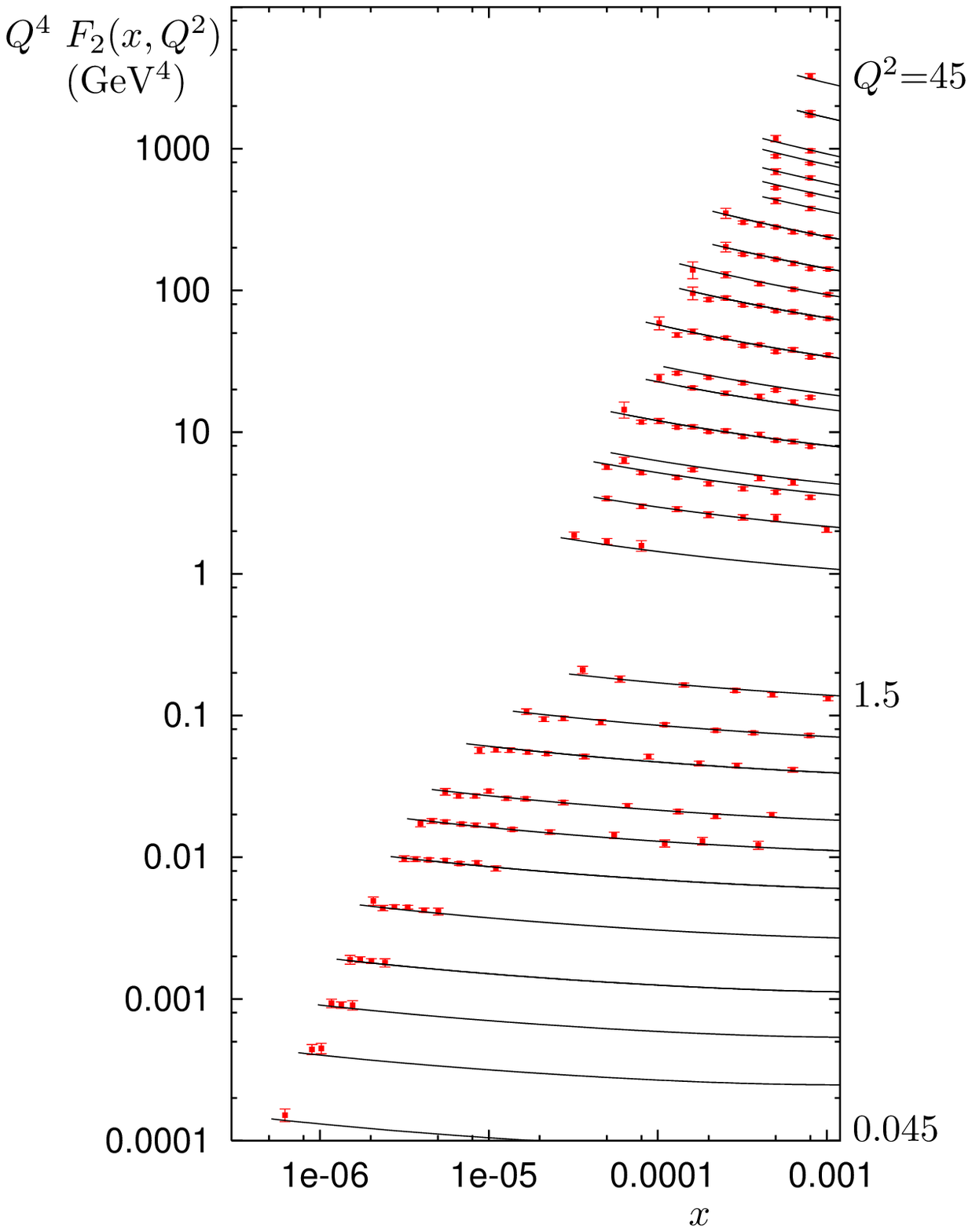}}

\centerline{(c)}

Figure 2: Fits using hard-pomeron, soft-pomeron and reggeon
exchange to total cross sections (the lower curves in each
plot are the hard-pomeron term) and to HERA data\defref\hera{
ZEUS collaboration: J Breitweg et al, Physics Letters B487 (2000) 53
and European Physical Journal C7 (1999) 609  and C21 (2001) 443\h
H1 collaboration: C Adloff et al, Physics Letters B520 (2001) 183
}
for the proton structure function $F_2(x,Q^2)$.
\endinsert

\topinsert
\vskip 10truemm
\centerline{\epsfxsize=0.54\hsize\epsfbox[53 500 480 760]{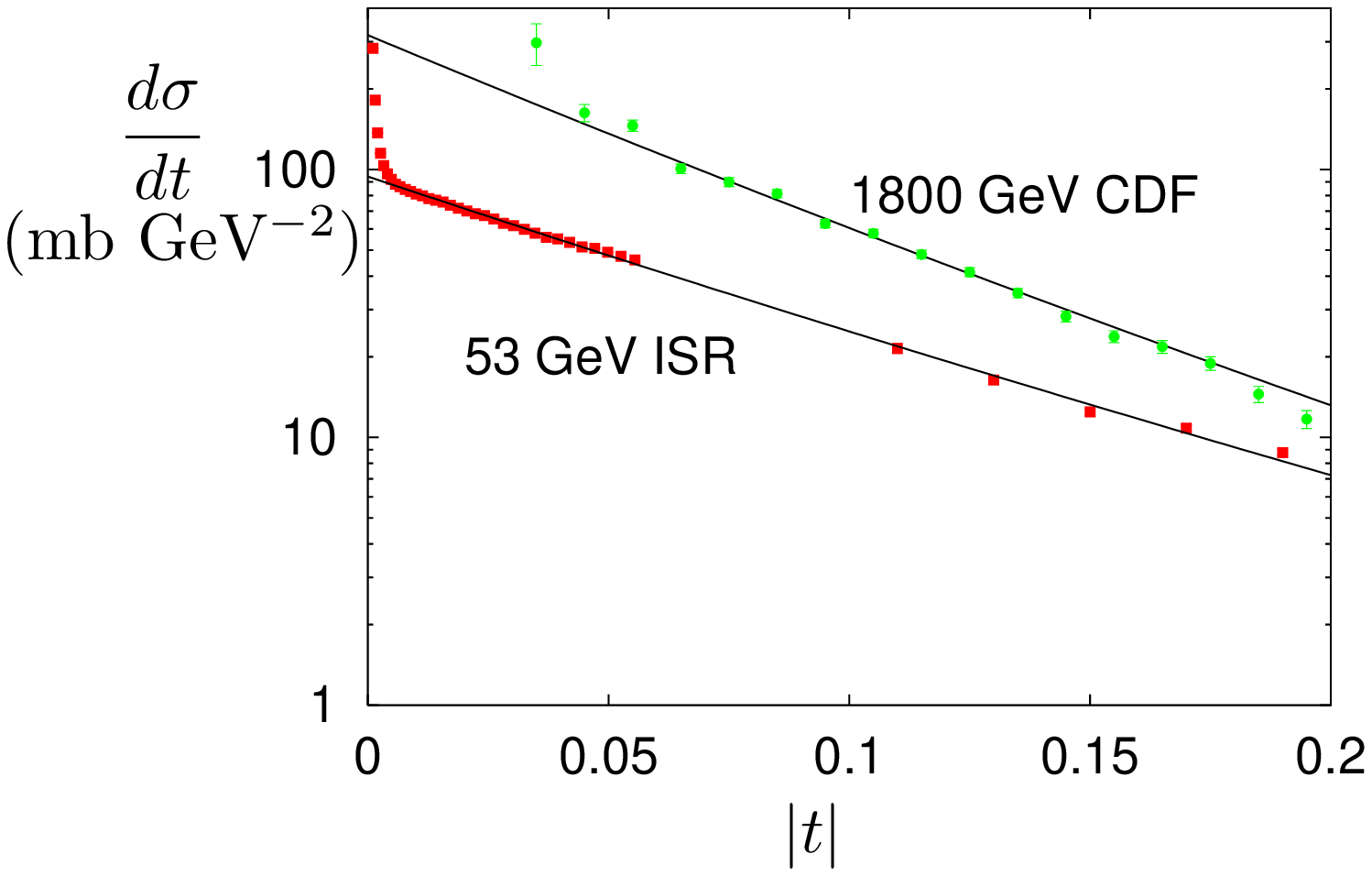}}

Figure 3: $pp$ elastic scattering at 53 GeV (CERN ISR\defref\chhav{
CHHAV collaboration: E Nagy et al, Nuclear Physics B150 (1979) 221
}) 
and $p \bar p$ at 1800 GeV (CDF\ref{\tevatron}), with the Regge-theory curves.
\vskip 10truemm
\centerline{\epsfxsize=0.55\hsize\epsfbox[120 585 360 770]{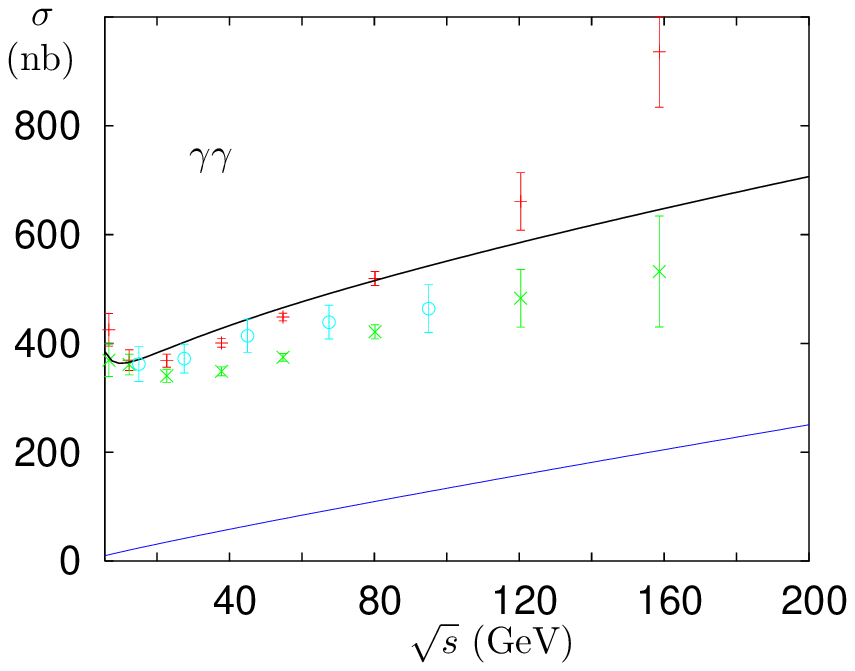}}

Figure 4: L3 data\defref\lthree{
L3 collaboration: Physics Letters B519 (2001) 33
}
for $\sigma^{\gamma\gamma}$ using two different Monte Carlos (black points),
and OPAL data\defref\opaltot{
G Abbiendi et al, European Physical Journal C14 (2000) 199
}
(open points),
together with curve obtained from Regge factorisation plus the box graph.
The lower curve is the hard-pomeron term.
\endinsert

\topinsert
\line{\hfill
\epsfxsize=0.34\hsize\epsfbox[85 600 295 765]{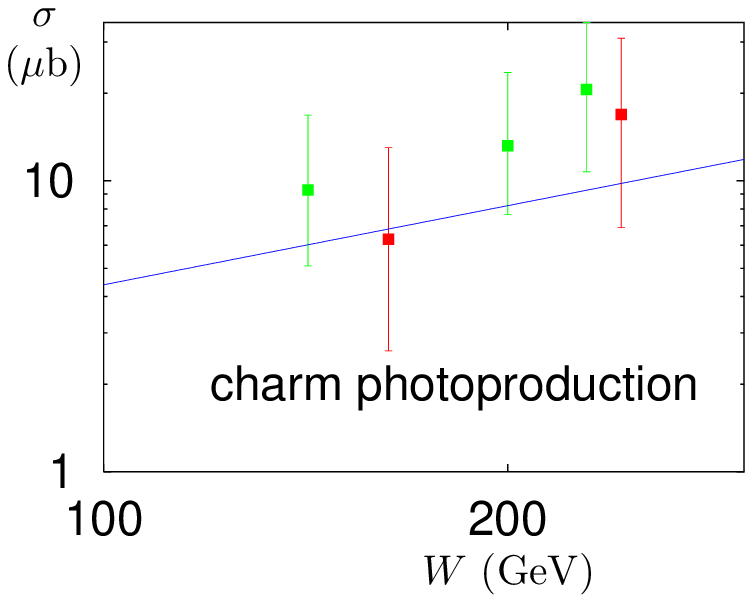}\hfill
\epsfxsize=0.5\hsize\epsfbox[65 295 377 755]{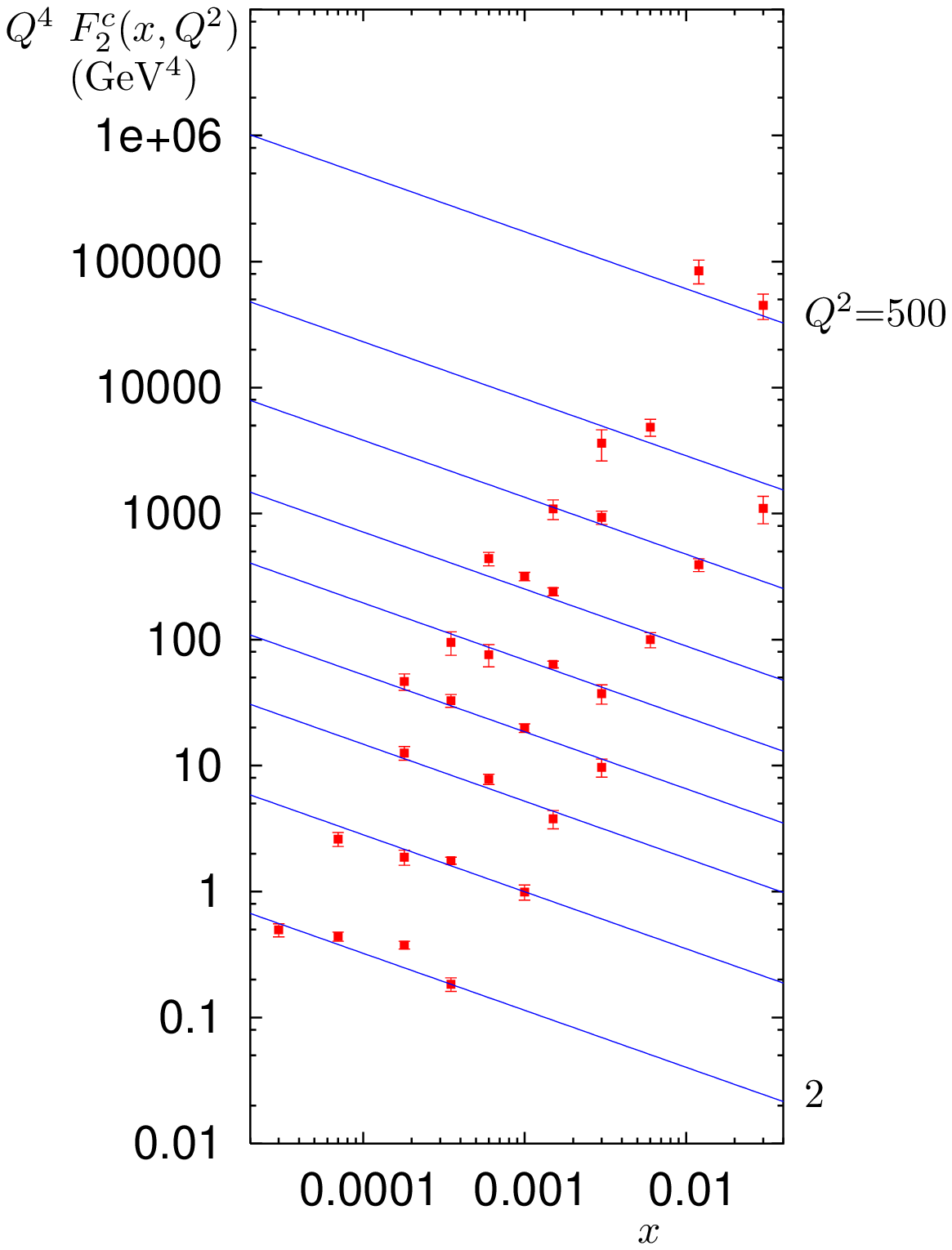}\hfill}

\line{\hfill (a)\hfill\hfill (b)\hfill}

Figure 5: The charm photoproduction cross section, 
and ZEUS data\defref\zeuscharm{ 
ZEUS collaboration: S Chekanov et al, Nuclear Physics B672 (2003) 3
}
for the charm structure function $F_2^c(x,Q^2)$ and 
The curves
are 0.4 times the hard-pomeron contributions to the curves in figures
2a and 2c
\vskip 10truemm
\centerline{\epsfxsize=0.5\hsize\epsfbox[110 500 485 775]{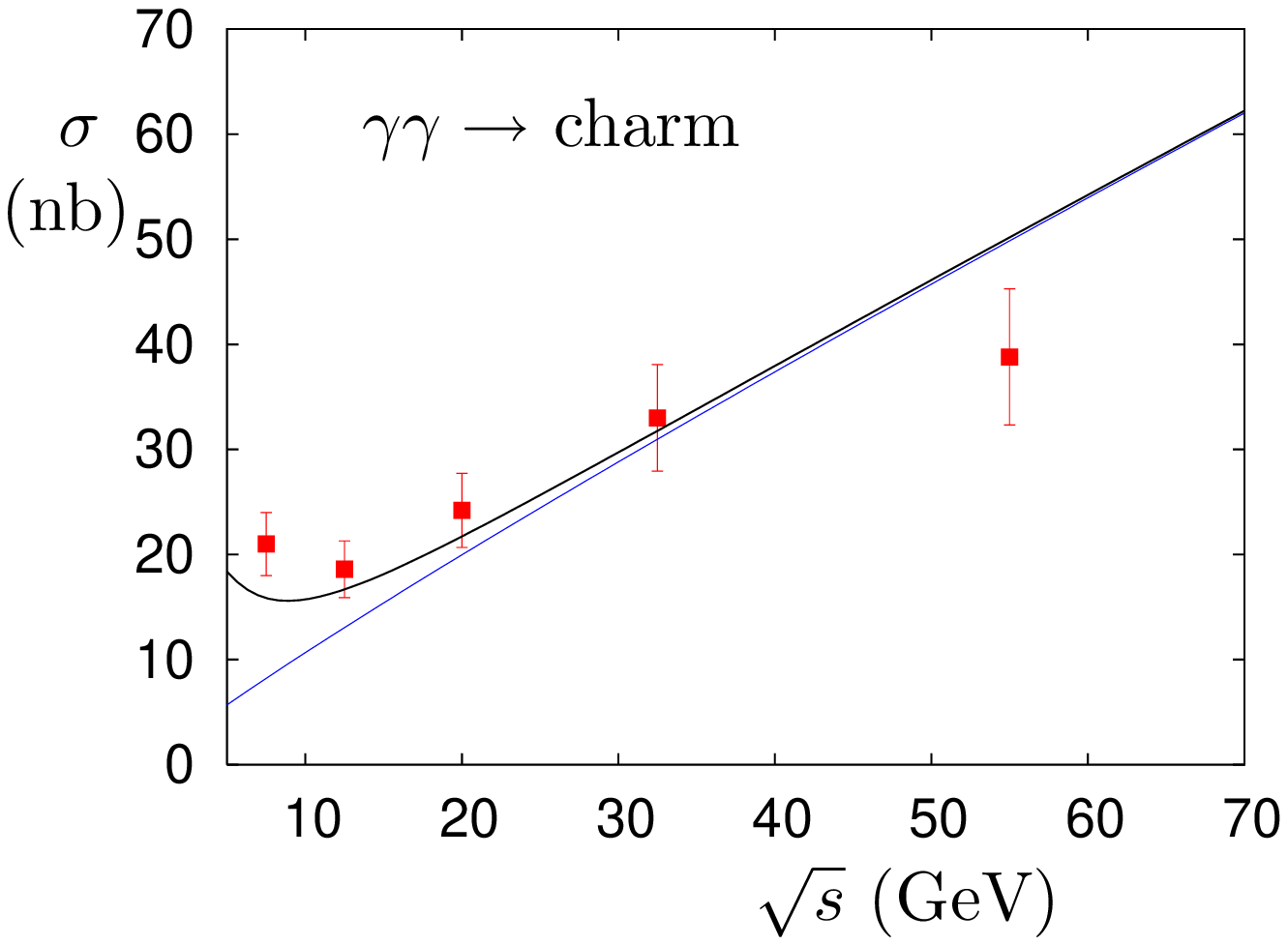}}

Figure 6: L3 data\defref\lthreech{
L3 collaboration: M Acciari et al, Physics Letters B514 (2001) 19
}
for $\gamma\gamma\to$ charm. The lower curve is the hard-pomeron contribution
obtained from Regge factorisation; the upper curve includes also the box
graph.
\endinsert
\topinsert
\line{
\epsfxsize=0.4\hsize\epsfbox[80 60 385 290]{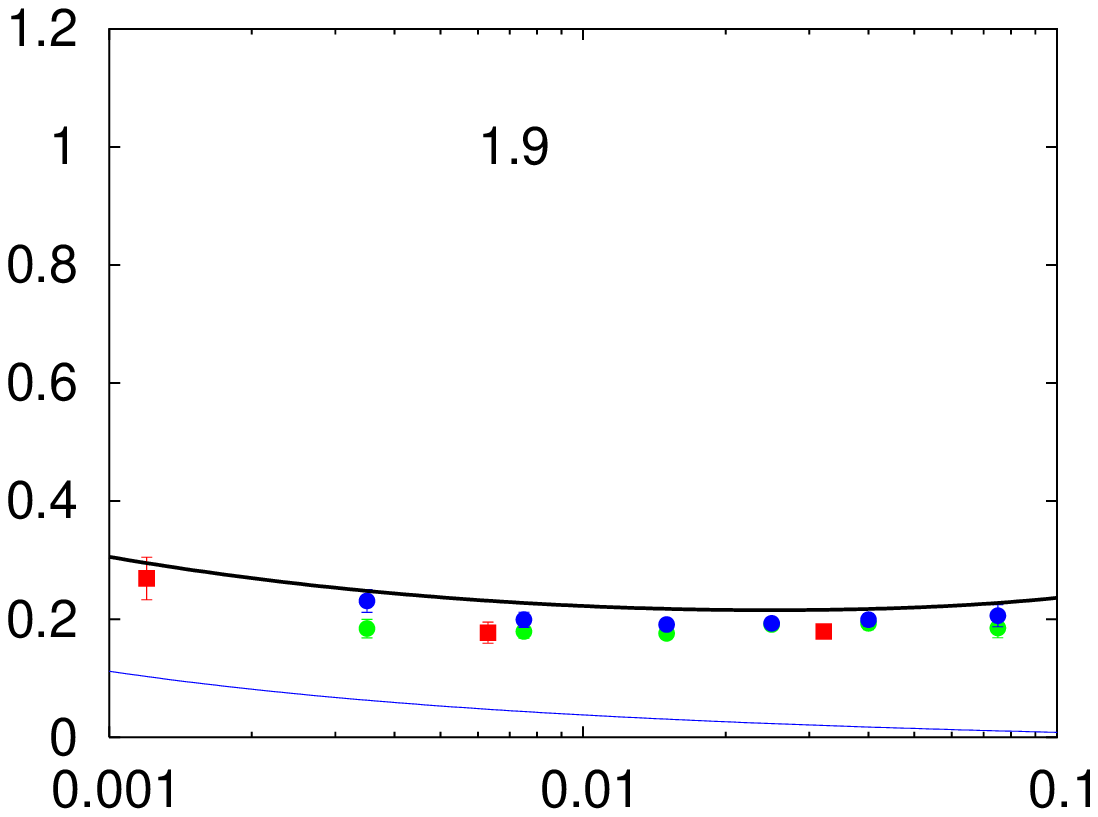}\hfill
\epsfxsize=0.4\hsize\epsfbox[80 60 385 290]{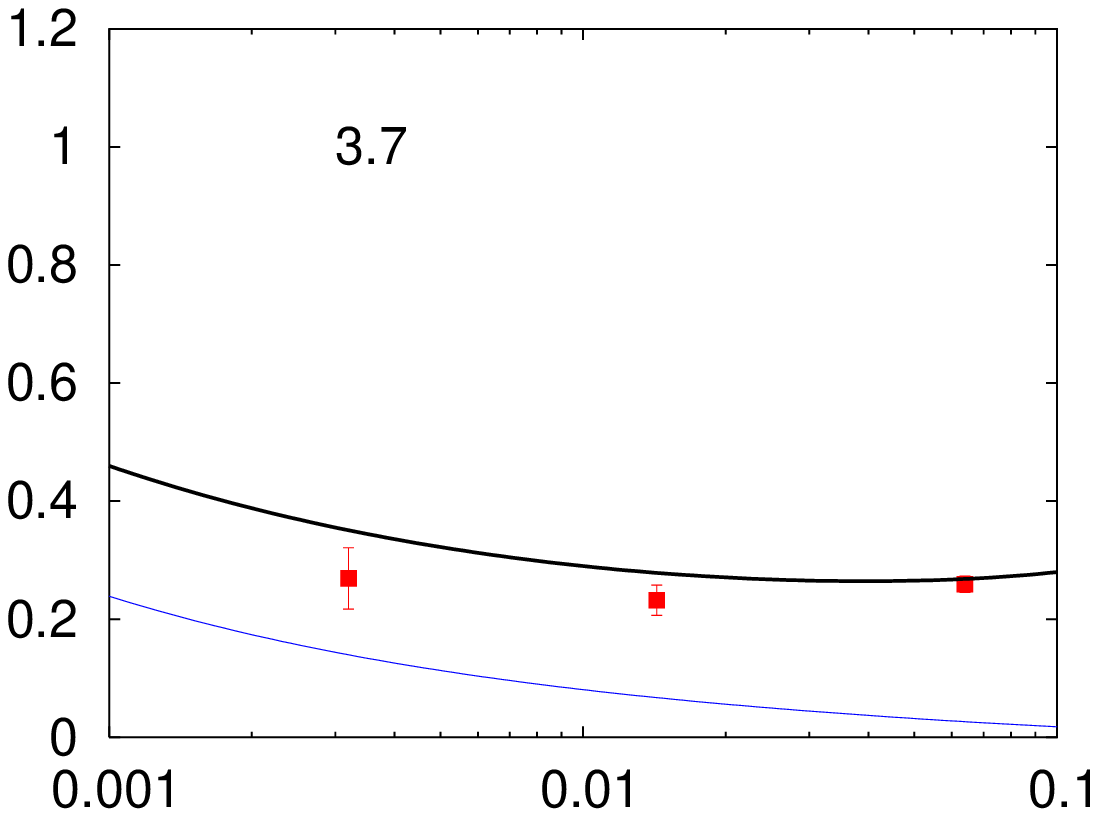}}
\vskip 3truemm
\line{
\epsfxsize=0.4\hsize\epsfbox[80 60 385 290]{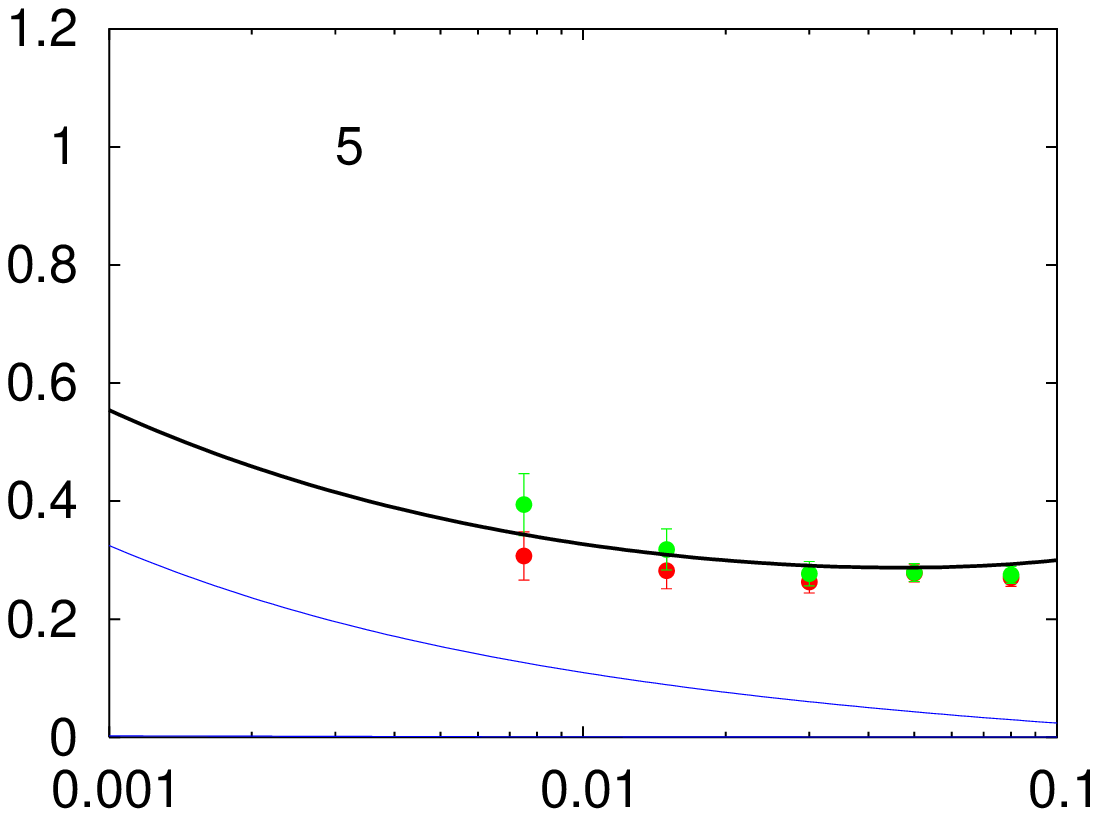}\hfill
\epsfxsize=0.4\hsize\epsfbox[80 60 385 290]{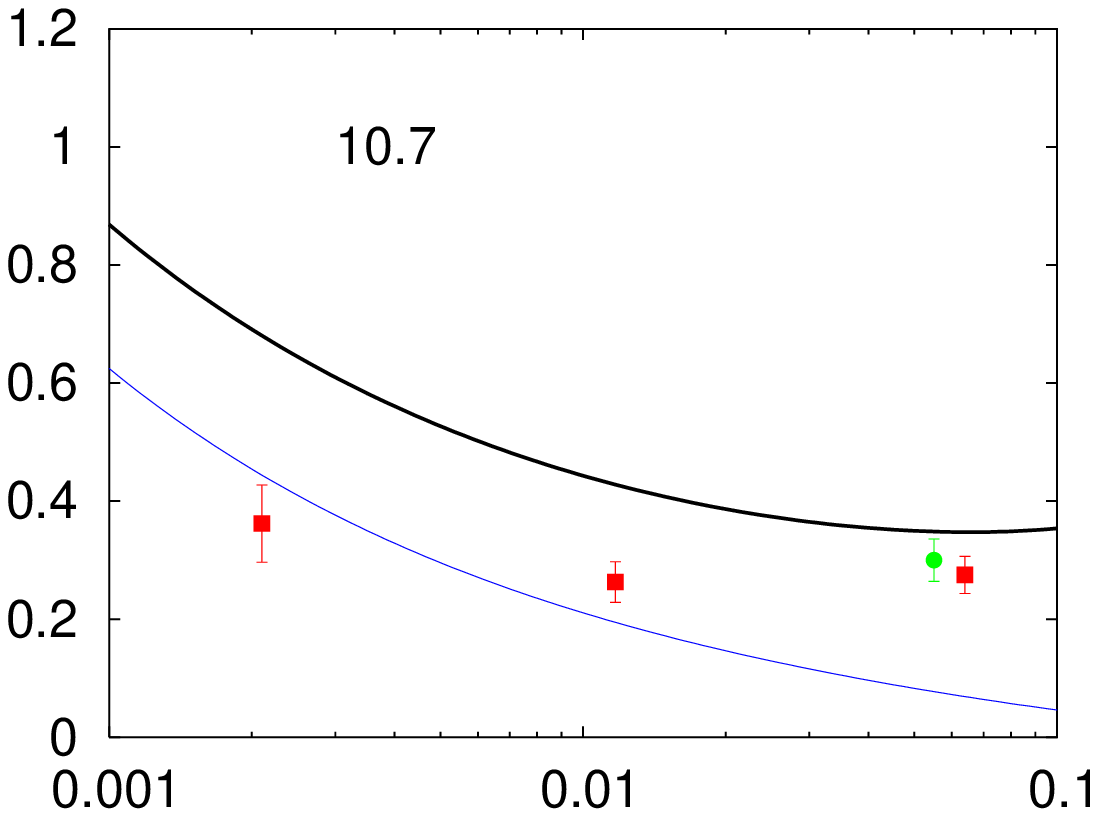}}
\vskip 3truemm
\line{
\epsfxsize=0.4\hsize\epsfbox[80 60 385 290]{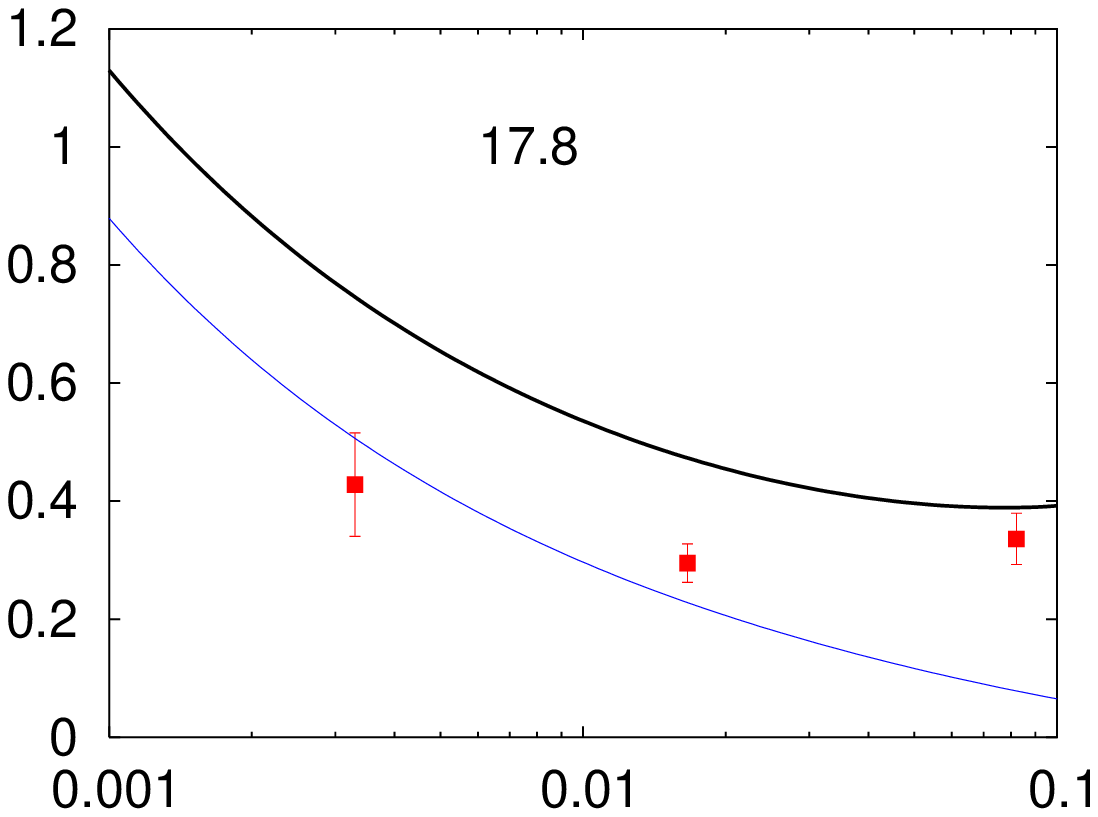}\hfill
\epsfxsize=0.4\hsize\epsfbox[80 60 385 290]{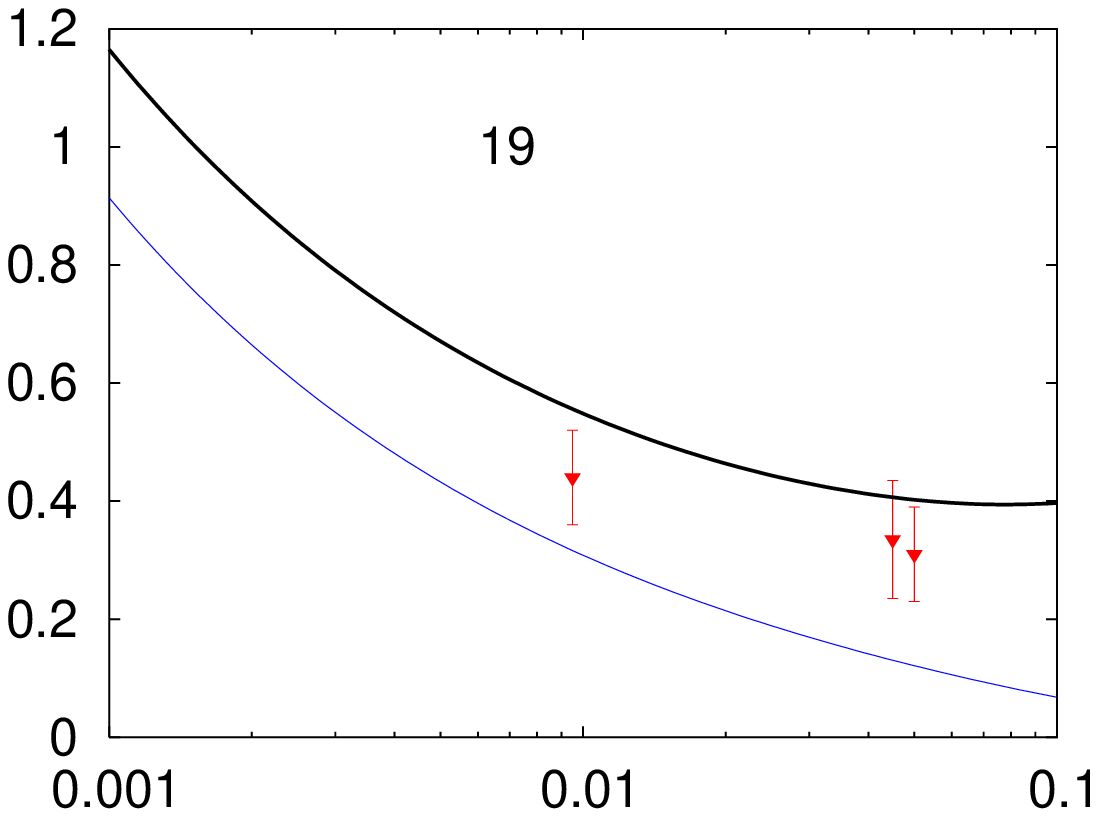}}
\vskip 3truemm
\line{
\epsfxsize=0.4\hsize\epsfbox[80 60 385 290]{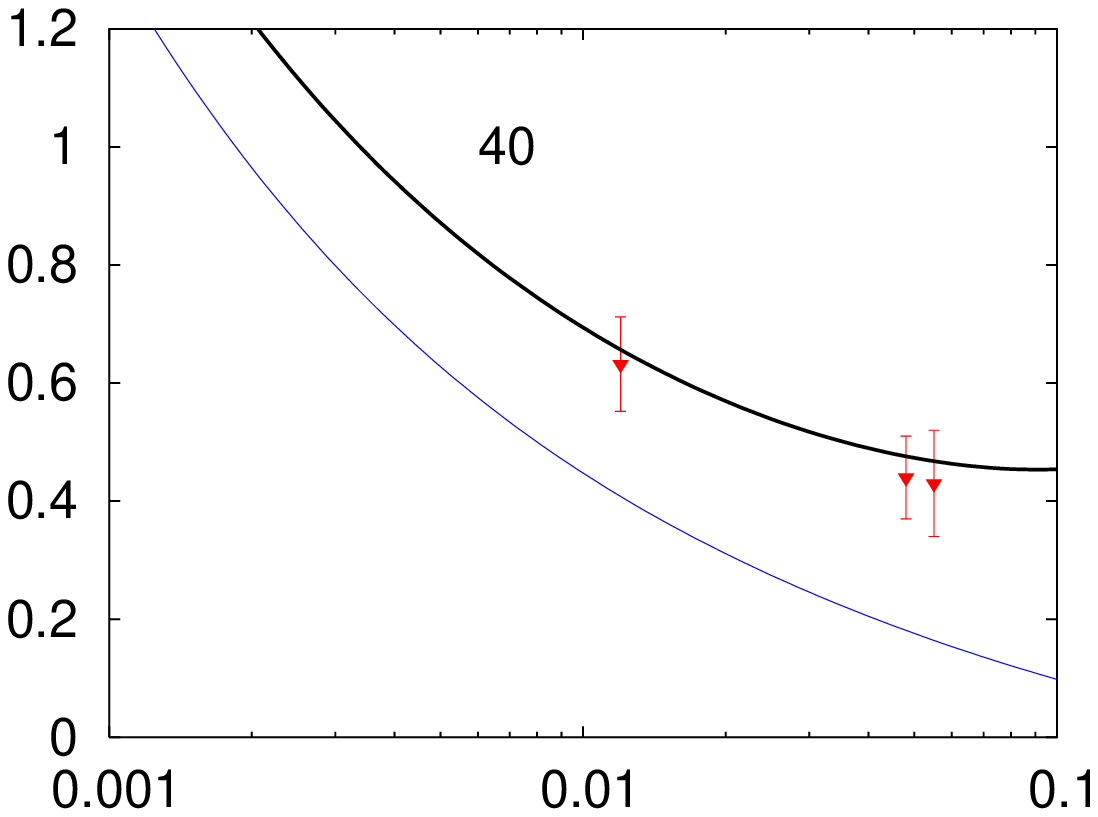}\hfill
\epsfxsize=0.4\hsize\epsfbox[80 60 385 290]{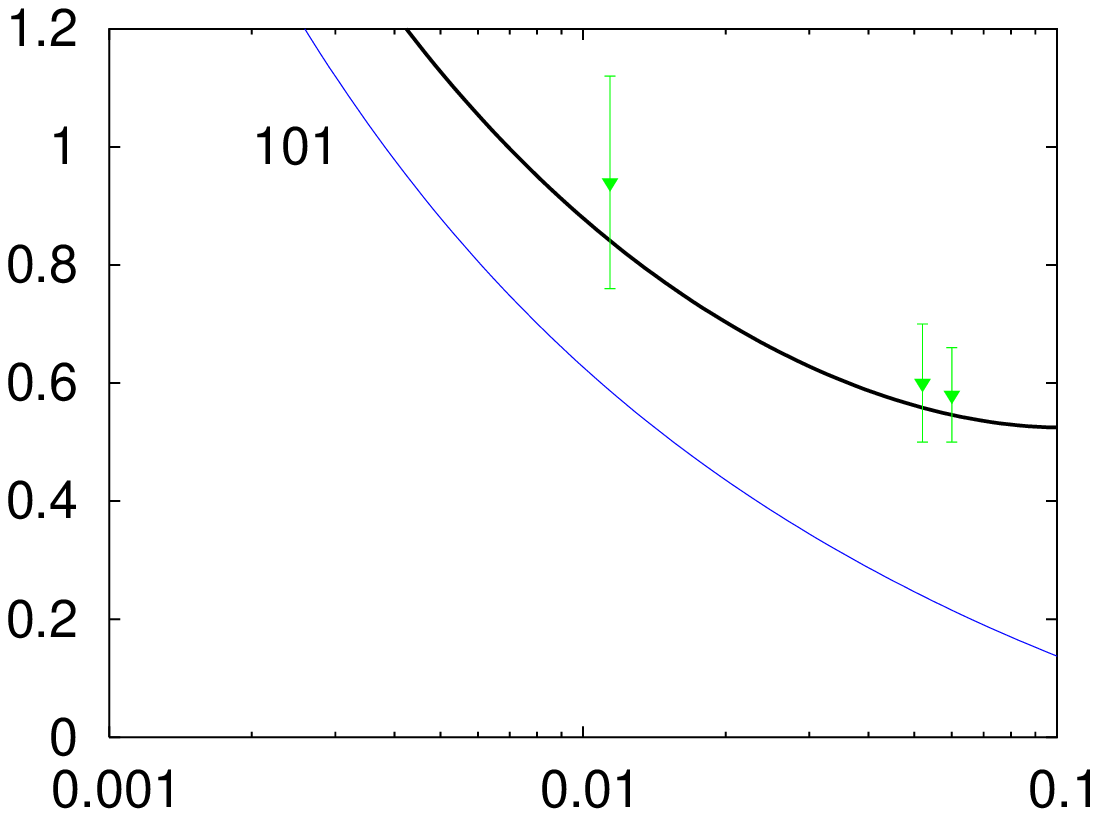}}

Figure 7: Data\defref\photstruct{
L3 collaboration: M Acciari et al, Physics Letters B436 (1998) 403\h
OPAL collaboration: G Abbiendi et al. 2000 European Physical Journal C18 (200) 15;
DELPHI collaboration, internal note 2003-025-CONF-645, available on\h
delphiwww.cern.ch/~pubxx/delsec/conferences/summer03/
}
the photon structure function $F_2^{\gamma}/\alpha$
plotted against $x$ at various values
of $Q^2$ with fits from Regge factorisation. In each case the lower curve is
the hard-pomeron contribution.
\endinsert
\topinsert
\centerline{\epsfxsize=0.42\hsize\epsfbox[125 500 470 775]{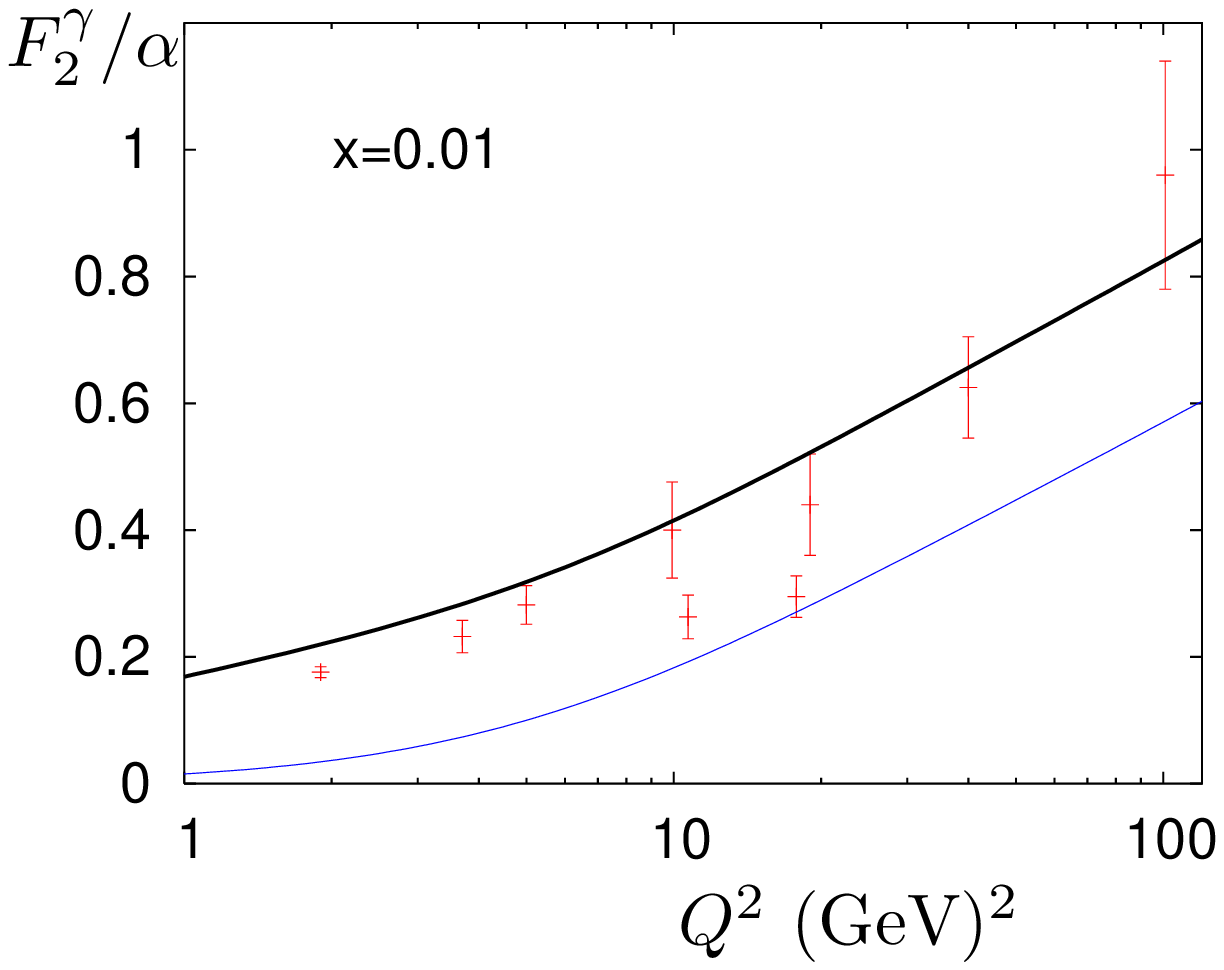}}

Figure 8: Data\ref{\photstruct}
for the photon structure function $F_2^{\gamma}/\alpha$ at $x$ close to 0.01
plotted against $Q^2$, with the prediction from Regge factorisation.
The lower curve corresponds to hard-pomeron exchange alone.
\vskip 10truemm
\centerline{\epsfxsize=0.42\hsize\epsfbox[100 585 355 775]{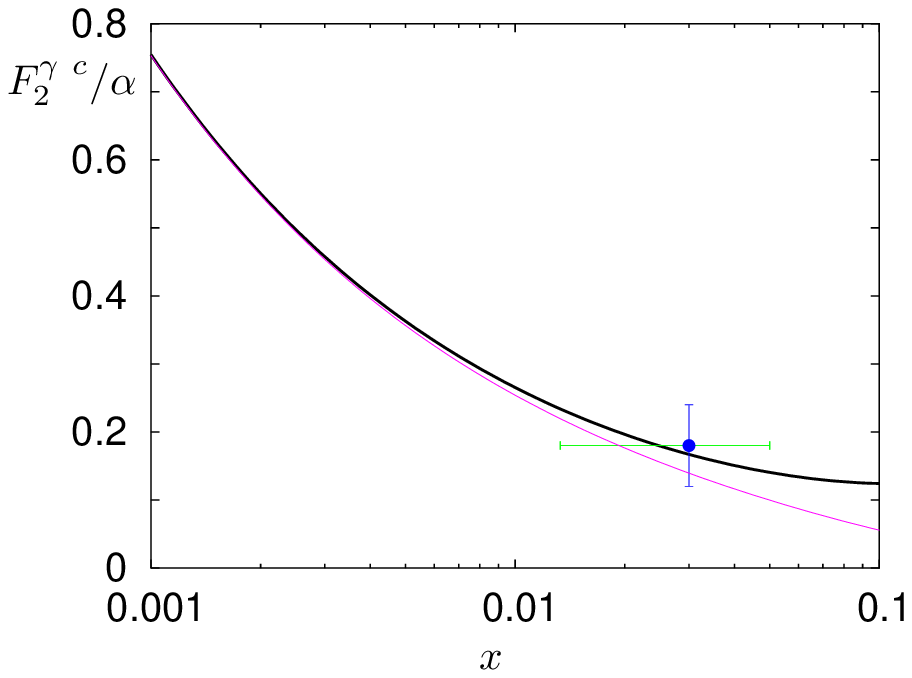}}

Figure 9: Data point\defref\opalch{
OPAL collaboration, G Abbiendi et al, Physics Letters B539 (2002) 13
}
for the photon structure function $F_2^{\gamma~c}/\alpha$ at $Q^2=20$ GeV$^2$
plotted against $Q^2$, with the prediction from Regge factorisation.
The lower curve corresponds to hard-pomeron exchange alone.
\vskip 10truemm
\centerline{\epsfxsize=0.42\hsize\epsfbox[110 585 355 775]{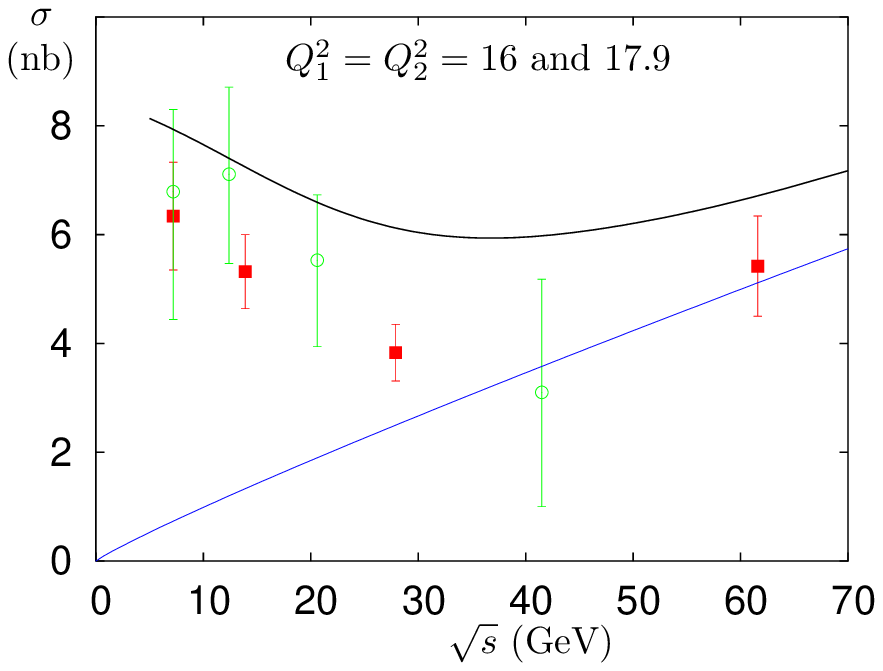}}

Figure 10: L3 and OPAL data\defref\gstar{
L3 collaboration: M Acciari et al, Physics Letters B514 (2001) 19\h
OPAL collaboration: G Abbiendi et al, European Physical Journal C24 (2002) 17
}
for $\sigma(\gamma^*\gamma^*)(W)$ at $Q^2=16$ and 17.9 Gev$^2$.
The lower curve corresponds to hard-pomeron exchange alone.
\endinsert
\topinsert
\centerline{\epsfxsize=0.42\hsize\epsfbox[115 495 460 765]{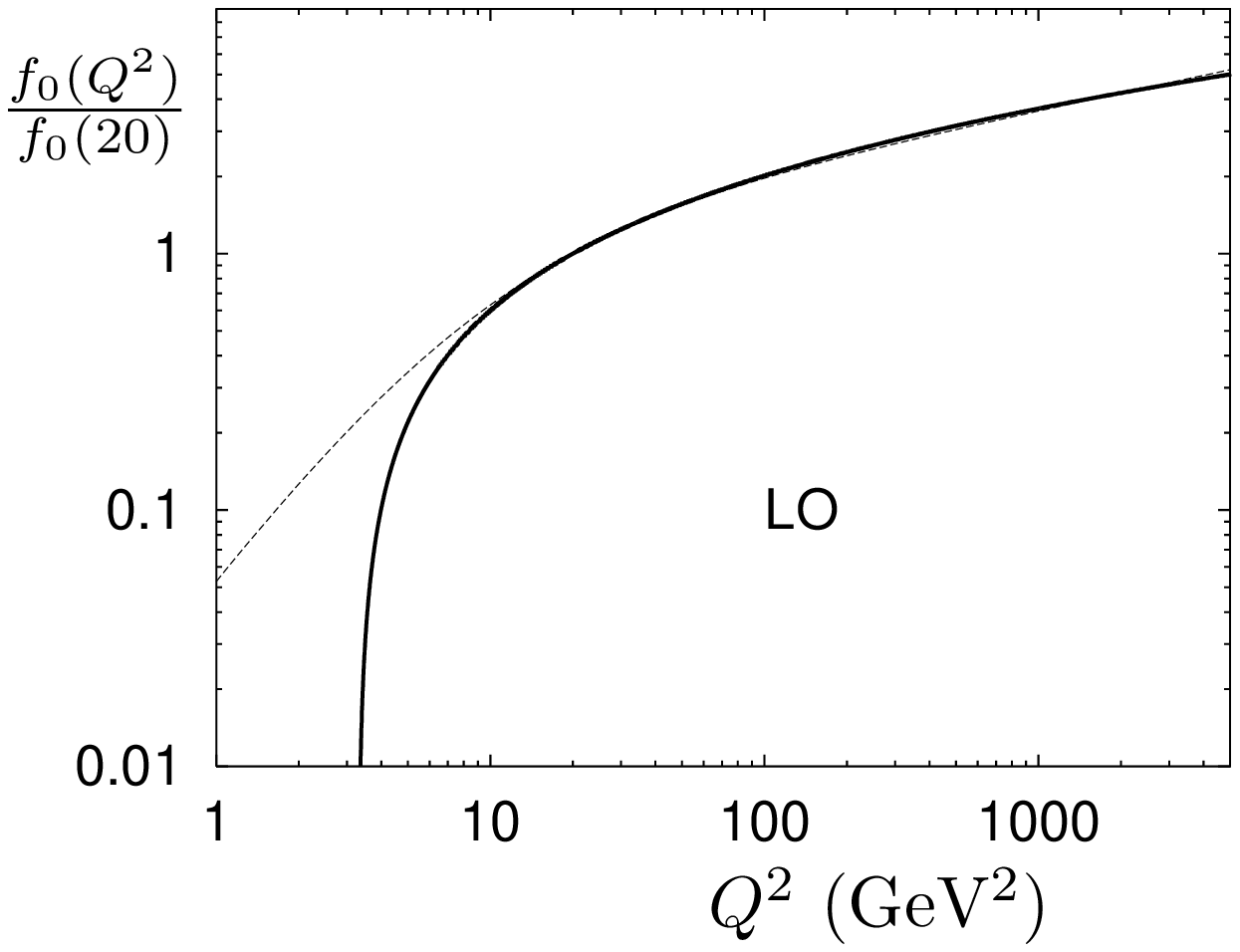}}

Figure 10: $Q^2$ evolution of the hard-pomeron coefficient function.
The upper curve is extracted from the data, and the lower corresponds to
LO DGLAP evolution. 
\endinsert

We perform a minimum-$\chi^2$ fit to the data for $\sigma^{pp},\sigma^{p\bar p},
\sigma^{\gamma p}$ and $F_2(x,Q^2)$.
(We have not included in
our fits the data for $\rho$, the ratio of the real and imaginary parts
of the forward hadronic amplitudes, because of the theoretical uncertainties
in the analysis needed to extract them from experiment.) 
This leads to the set of parameters
$$
\epsilon_0=0.452~~~~~~~~\epsilon_1=0.0667~~~~~~~~\epsilon_R=-0.476
$$$$
X_0^{pp}=X_0^{p\bar p}=0.0139~~~~~~~X_1^{pp}=X_1^{p\bar p}=24.22
$$$$
Y^{pp}=46.55~~~~~~~Y^{p\bar p}=95.81
$$$$
A_0=0.00151~~~~Q_0^2=7.85~~~~A_1=0.658~~~~Q_1^2=0.600~~~~A_R=1.01~~~~Q_R^2=0.214
\eqno(5a)
$$
so that
$$
X_0^{\gamma p}=0.000169~~~~X_1^{\gamma p}=0.0737~~~~Y^{\gamma p}=0.113 
\eqno(5b)
$$

The resulting fits to the $\gamma p,~pp$ and $p\bar p$ total cross sections
are shown in figures 2a and 2b. They do not look all that different from the
old fits without the hard pomeron. Together, the 
$\epsilon_0=0.45$  hard pomeron and the $\epsilon_1=0.067$ soft pomeron
behave not very differently from a single $\epsilon=0.08$ pomeron.
We may check this by considering $pp$ and $p\bar p$ elastic scattering.
We introduce the Regge trajectories
$$
\alpha_0(t)=1+\epsilon_0+\alpha'_0t
$$$$
\alpha_1(t)=1+\epsilon_1+\alpha'_1t
$$$$
\alpha_R(t)=1+\epsilon_R+\alpha'_Rt
\eqno(6)
$$
There is evidence from $J/\psi$ photoproduction that $\alpha'_0$
is small\defref\vector{
A Donnachie and P V Landshoff, Physics Letters B478 (2000) 146
}.
If $\alpha_0(t)$ and $\alpha_1(t)$ together resemble the single
canonical trajectory with slope $\alpha'=0.25$ (in GeV units), we
need $\alpha'_1$ to be greater than 0.25. We choose
$$
\alpha'_0=0.1~~~~~~~~~~~~~~\alpha'_1=0.3
\eqno(7)
$$
though these values are open to some adjustment. The resulting
differential cross sections are shown in figure 3. Note that we have plotted
the CDF data at $\sqrt s=1800$ GeV rather than E710, because our fit
to the total cross section favoured CDF rather than E710. As in the 
past\defref\elastic{
A Donnachie and P V Landshoff, Nuclear Physics B231 (1984) 189
}\ref{\book},
we have coupled the pomerons to the proton through the Dirac elastic
form factor $F_1(t)$.

If we apply the factorisation (1), apply it similarly to
soft-pomeron and regge exchange, and add in a contribution from
the simple box graph, we obtain the $\gamma \gamma$ total cross section shown 
in figure 4. The data need large acceptance corrections
and so are highly sensitive to the Monte Carlo used for this.
The L3 data\ref{\lthree} shown are the outputs from two of their Monte Carlos.

Regge factorisation should apply also to cross sections for charm production. 
We have already observed\ref{\twopom}\defref\charm{
A Donnachie and P V Landshoff, Physics Letters B550 (2002) 160
}
that the data for the charm structure function\ref{\zeuscharm} $F_2^c$
show that at small $x$ it is described by hard-pomeron exchange alone.
Further, it is given by 0.4 times the hard-pomeron component of  $F_2$. 
That is,
the hard pomeron's couplings to $u,d,s,c$ quarks are equal. This
remains true even down to $Q^2=0$; see figure 5. 
Combining the flavour-blindness of the hard-pomeron coupling with
Regge factorisation, we conclude that the hard-pomeron contribution
to the cross section for \hbox{$\gamma\gamma\to$ charm} should
be $0.64~[\sigma_c^{\gamma p}]^2/\sigma_{{\fiverm{\hbox{HARD}}}}^{pp}$.
This corresponds to the lower curve in figure 6. Adding in the box graph,
with $m_c=1.3$ GeV, gives the upper curve.

As we have said, the factorisation should apply also at nonzero $Q^2$.
We apply it as in (2), and treat similarly soft-pomeron and reggeon
exchange. Ideally, we should keep to $x<0.001$, as in figure 2c, as for larger
$x$ we need to include some factor of $x$ which is equal to 1 at small
$x$ but $\to 0$ as $x\to 1$. This factor is unknown, so we simply
use a power of $(1-x)$ determined by the old
dimensional counting rules\defref\dimcount{
S~J Brodsky and G~R Farrar, Physical Review Letters  31 (1973) 1153\h
V~A Matveev, R~M Murddyan and A~N Tavkhelidze, Lettere 
Nuovo Cimento 7 (1973) 719
}.
This leads us to include a factor $(1-x)^5$ in each of the pomeron
contributions, and $(1-x)$ in the reggeon contribution. These factors 
are not correct, as they take no account of perturbative evolution,
but it is better to include them than not to do so. Adding in
the box graph gives us the results for the photon structure
function $F_2^{\gamma}(x,Q^2)$ shown in figure 7. 
Again, in each case the lower curve is the hard-pomeron component.
It seems that the OPAL data at $Q^2=10.7$ and 17.8 GeV$^2$ may be out
of line with the other data. This impression is enhanced if we plot data
at values of $x$ close to 0.01 against $Q^2$, as in figure 8.
Figure 9 shows the prediction for the photon's charm structure function
at $Q^2=20$ GeV$^2$, together with the data point from OPAL\ref{\opalch}.

Lastly, we consider the cross section for $\gamma\gamma\to$ hadrons
when both photons are off shell: see figure 9.

We are not able to reach a firm conclusion about whether the hard pomeron
obeys Regge factorisation, because the tests involve data that require large
Monte Carlo acceptance corrections and are therefore subject to no little
uncertainty. We have applied several tests: $\sigma^{\gamma\gamma},
\sigma^{\gamma\gamma}_c, F_2^{\gamma}$ and $\sigma^{\gamma^*\gamma^*}$.
None of them clearly fails.

Note that in any case factorisation should not be exact. The hard and
soft pomeron contributions are not really powers of $s$ or $1/x$. The
powers we have used should be regarded as effective powers. Assuming,
as we have, that each pomeron yields a simple power $e_0$ or $e_1$,
then there are corrections from the exchange of two or more pomerons.
These are negative and so result in effective powers $\epsilon_0<e_0$
and $\epsilon_1<e_1$. If factorisation is approximately satisfied,
this is an indication that the corrections are small, so that $e_0$ is
only a little greater than $\epsilon_0$ and $e_1$ is a little greater
than $\epsilon_1$. When we analysed\defref\elastic{
A Donnachie and P V Landshoff, Nuclear Physics B267 (1986)  690
} 
$pp$ elastic scattering data using only the soft pomeron, we concluded
that the correction term was at the 10\% level for the total cross
section at Tevatron energy. It will surely be higher at LHC energy,
so that the hard-pomeron effective power will be reduced at this
energy. Without such a reduction, the cross section at LHC energy
would extrapolate to 165 mb. So, while we do not believe that it
will be as large as this,  there is a real prospect that the
cross section will be found to be significantly larger than the
prediction of 101 mb given by our old data fit\ref{\sigtot} fit based
on the soft pomeron alone.

Related analyses to this one have been performed by Donnachie and 
Dosch\defref\dd{
A Donnachie and H G Dosch, 
Physical Review D65 (2002) 014019
}
and by Cudell and collaborators\defref\cudellfac{
J R Cudell, E Martynov, O Selyugin and A Lengyel, hep-ph/0310198
},
though the details are very different. 
The former is based on a dipole picture, but gives results similar to 
ours, while the latter maintains that corrections from multiple exchanges 
are large.

Finally, we note that, as with our previous analysis\defref\evol{
A Donnachie and P V Landshoff, Physics Letters B533 (2002) 277
and B550 (2002) 160 
},
as soon as $Q^2$ is large enough for the DGLAP equation to be valid,
the variation of the hard-pomeron coefficient function $f_0(Q^2)$
we have extracted from the data and given in (4) and (5),
agrees exactly with the predictions of LO evolution. 
We show this in figure 10. As before\ref{\evol}, NLO evolution gives
almost the same result.

\vskip 13truemm
{\sl This research was supported in part by PPARC}
\vfill\eject
{\medskip\immediate\closeout\rfile\writestoppt
\baselineskip=7pt{{\bf References}}\bigskip{\frenchspacing%
\parindent=20pt\escapechar=` \input refs.tmp\bigskip}\nonfrenchspacing}

\bye